\documentclass[twocolumn,showpacs,preprintnumbers,amsmath,amssymb]{revtex4}

\usepackage{graphicx}
\usepackage{dcolumn}
\usepackage{bm}

\begin{document}


\preprint{3P Methodology}

\title{ Methodology for the study of modified jet-like topologies in heavy ion collisions \\ via three particle correlation functions 
 
}

\author{ N.~N.~Ajitanand} 
\author{ J.~M.~Alexander}
\author{ Roy~A.~Lacey}
\author{A.~Taranenko}
%
\affiliation{Department of Chemistry, 
Stony Brook University, \\
Stony Brook, NY, 11794-3400, USA}

\date{\today}

\begin{abstract}
Methodology is presented for analysis of three-particle correlation 
functions obtained in heavy ion collisions at ultra-relativistic energies.  
We show that harmonic correlations can be removed and jet driven 
correlations reliably extracted.  Results from detailed Monte Carlo simulations 
are used to demonstrate the efficacy of this technique for the study of 
modifications to away-side jet topologies. Such modifications are an essential probe 
of the properties of the quark gluon plasma produced in heavy ion collisions.
\end{abstract}

\maketitle

\section{\label{sec:level1}Introduction}

Studies of ultra-relativistic heavy ion collisions have provided a 
wealth of evidence for the creation of a new state of matter in a collision zone of very 
high energy density \cite{Adams:2005dq,Adcox:2004mh,Arsene:2004fa,Back:2004je}, 
historically termed the quark gluon plasma QGP. 
As the properties of this matter are being explored and 
characterized, it may be that new terminology will be chosen to be more 
suggestive of its actual properties, as they are revealed.  Current 
research is actively directed toward measurements of these properties.  

	One of the major tools for such measurements has been the study of 
correlations between the observed particles which emanate from 
the collision medium following its expansion and ultimate hadronization. 
For example, much detailed information has been obtained for the so called ``harmonic 
flow'' correlations between individually selected particles and the 
reaction plane (e.g. the second harmonic coefficient  ($v_2$)) as a 
function of particle identity PID, collision centrality, transverse 
momentum $p_T$, and rapidity $\eta$ \cite{Adcox:2002ms,Voloshin:2002wa,Adams:2003zg,
Adler:2003kt,Adams:2004bi,Alver:2006wh,Adler:2004cj,Afanasiev:2007tv,Afanasiev:2009wq,
Lacey:2009xx}.  From the set of systematic data, initial estimates have been made of 
the speed of sound in the QGP, the ratio of its viscosity to entropy density, its 
bulk viscosity and other properties \cite{Shuryak:2004cy,Lacey:2006pn,Lacey:2006bc,Drescher:2007cd,Xu:2007jv,
Song:2008si,Greco:2008fs,Lacey:2009xx,Chaudhuri:2009ud}.

	Formation of the collision medium is sometimes accompanied by hard parton-parton 
scatterings. These scattered partons can interact strongly with the medium and 
lose energy as they propagate through it, before fragmenting  into 
jets of hadrons~\cite{Baier:1996kr,Gyulassy:2003mc,Kovner:2003zj}. 
Such energy loss can lead to a strong modification of both the yield and 
the topological patterns of jets~\cite{Adler:2005ee,Adare:2007vu}. This 
gives another detailed experimental probe of the medium which is 
currently being explored and developed via measurements of di-jet 
correlations 
\cite{Adler:2005ee,Adams:2005ph,Adare:2006nr,Ajitanand:2006is,Adare:2007vu,Abelev:2008nda}. 
%
%

Techniques for the study of jet-induced two-particle azimuthal angle correlations are well 
advanced, and results are being accumulated for systematic evaluation. 
In p+p and d+Au collisions, these  jets of hadrons are 
found essentially back-to-back in azimuth i.e. $\Delta\phi \sim 180^0$ \cite{Adler:2005ad}.
By contrast, the observed di-jet topologies in Au+Au collisions are found to be significantly 
modified \cite{Adler:2005ee,Adams:2005ph,Adare:2006nr,Ajitanand:2006is,
Adare:2007vu,Abelev:2008nda}, 
presumably due to parton medium interactions that occur on passage of the parton through the 
reaction medium. One very intriguing result of these di-jet studies is the observation 
of broadening of the away-side jet, and even the displacement of its most 
probable angle away from $\Delta\phi=180^0$ \cite{Adler:2005ee,Adare:2007vu}. 
Several mechanistic scenarios have been proposed for this observation; they 
include {$\breve{C}$erenkov} gluon 
radiation \cite{Koch:2005sx}, conical flow \cite{Stocker:2005,CasalderreySolana:2004qm,
Renk:2005si,Betz:2008js,Neufeld:2008fi,Chesler:2007an} and deflected 
or ``bent'' jets \cite{Armesto:2004pt,Chiu:2006pu}. To date, the characteristic $p_T$ dependent 
patterns predicted for {$\breve{C}$erenkov} gluon radiation \cite{Koch:2005sx} have not been 
observed. The confirmation of a conical flow signal would not only give a direct probe of the equation 
of state (EOS) of hot QCD matter, but also an important constraint for 
an upper limit for the viscosity of the medium \cite{Stocker:2005,CasalderreySolana:2004qm,
Neufeld:2008fi,Chesler:2007an}.

%
%

Two particle correlation measurements do not provide an unambiguous distinction 
between conical flow and deflected jets. However, the topological information
afforded by the correlations between three or more particles can.  
Here, we lay out a method of analysis for three particle correlations which 
demonstrates a topological distinction between conical flow and deflected
(or bent) jets. We follow and build on methodology formerly presented for the study of 
two-particle correlations; namely, the use of normalized correlation 
functions and extensive testing via Monte Carlo reaction 
simulations \cite{Ajitanand:2005jj}.  To focus on the di-jet-like characteristics, 
we exploit a novel coordinate system that is most intuitive and natural for 
visualizing di-jet topologies. 
 
\section{\label{sec:level1}Two-particle correlation functions}

	In reference \cite{Ajitanand:2005jj} we presented a method for constructing 
and analyzing two particle correlation functions based on the relative laboratory 
azimuthal angle $\Delta\phi$, for particle pairs. In brief, jet correlations were emphasized by 
selecting events with at least one high transverse momentum $p_T$ (trigger) particle. 
Each trigger particle was then paired with associated particles of  
a lower $p_T$ to obtain the pair correlation function $C_2(\Delta\phi)$; 
\begin{eqnarray}
C_2(\Delta\phi) = \frac{N_{R}(\Delta\phi)}{N_{M}(\Delta\phi)},
\end{eqnarray}
where $N_{S}(\Delta\phi)$ and $N_{M}(\Delta\phi)$ are normalized same-events and 
mixed-events distributions and $\Delta\phi = |\phi_1-\phi_2|$ is the difference 
between the azimuthal angles of the particle pair.
The same-events distribution was constructed from particle pairs obtained from the same 
event; the mixed-events distribution was constructed by selecting each particle in a given 
pair from a different events having similar centrality and collision vertex positions.

\subsection{\label{sec:level1}Extraction of Jet Shapes}

	Using a two component model ansatz, we showed  \cite{Ajitanand:2005jj} that the pair
correlation from a combination of flow and jet sources is given by; 
\begin{eqnarray}
C_2(\Delta\phi) = b_0[C_H(\Delta\phi)+C_J(\Delta\phi)],
\label{C2}
\end{eqnarray}
where the flow contribution $C_H(\Delta\phi)$ can be estimated as
\begin{eqnarray}
C_H(\Delta\phi) = [1 + 2v_2^2cos2(\Delta\phi) + 2v_4^2cos4(\Delta\phi)],
\label{v2v42pc}
\end{eqnarray}
and $C_J(\Delta\phi)$ is the
jet function that needs to be evaluated.  It is noteworthy that no explicit 
or implicit assumption is made for the functional form of $C_J(\Delta\phi)$.
By rearrangement of Eq.~\ref{C2} one obtains
\begin{eqnarray}
C_J(\Delta\phi) =\frac{C_2(\Delta\phi)- b_0C_H(\Delta\phi)}{b_0}.
\label{J2}
\end{eqnarray}
Thus, knowledge of $b_0$ is required to evaluate the jet-like function $C_J(\Delta\phi)$. 
It is clear from Eq.~\ref{J2} that $b_0$ is influenced by the jet function which is 
being sought after, and an approximation is required to aid the evaluation of $b_0$. 
After extensive detailed simulation studies, we concluded that 
a reasonable assumption for the extraction of reliable jet-like functions 
is the zero yield at minimum (ZYAM) condition --  that the di-jet function  
has a zero yield at minimum \cite{Ajitanand:2005jj};
\begin{eqnarray}
b_0C_H(\Delta\phi_{min}) = C_2(\Delta\phi_{min}),
\end{eqnarray}
which can be solved to obtain $b_0$. This procedure allows reliable 
extraction of jet-like topologies and a lower limit for jet-like per 
trigger yields.

\subsection{\label{sec:level1}Simulations}

	Reaction simulations were used to extensively test this analysis approach. 
The requisite simulations were carried out on an event-by-event basis with the 
following essential steps: 
\begin{itemize}
  \item First, the reaction plane orientation was chosen.
  \item Particles were then emitted with $p_T$ and multiplicity according to the observed 
        distributions.
  \item For flowing particles, the azimuthal angle for each particle $\phi_i$ was chosen to give 
        a harmonic distribution with respect to the angle of the reaction plane $\psi_R$:
\begin{eqnarray}
N(\phi-\psi_R) \propto [1 + 2v_2cos2(\phi-\psi_R) \nonumber \\ + 2v_4cos4(\phi-\psi_R)]
\label{v2v4}
\end{eqnarray}
where $v_{2,4}$ are Fourier coefficients which characterize the strength of the flow. 
  \item For jets, the orientation of the lead- or near-side axis was chosen with a random azimuth. 
        For ``normal'' jets, the away-side axis was oriented opposite to the lead-axis on average.
  
  \item The lead-jet particles were emitted clustered about the lead-axis. The 
away-side jet particles were emitted clustered about the away-side axis.
\end{itemize}
Simulations were performed both for an ideal detector and for the PHENIX detector. 
A detailed description of the latter can be found in Ref.~\cite{Adcox:2003zm}.
As outlined in Ref.~ \cite{Ajitanand:2005jj}, the simulations indicated that, even for 
cases in which strongly distorted away-side jets were introduced, the decomposition method retrieved 
the shape of the input jet function in detail, confirming that the decomposition procedure is robust.

We now turn to the discussion of 
three particle correlation functions. 

\section{\label{sec:level1}Three-particle correlation functions}

\begin{figure*}[t]
\includegraphics[width=0.75\linewidth]{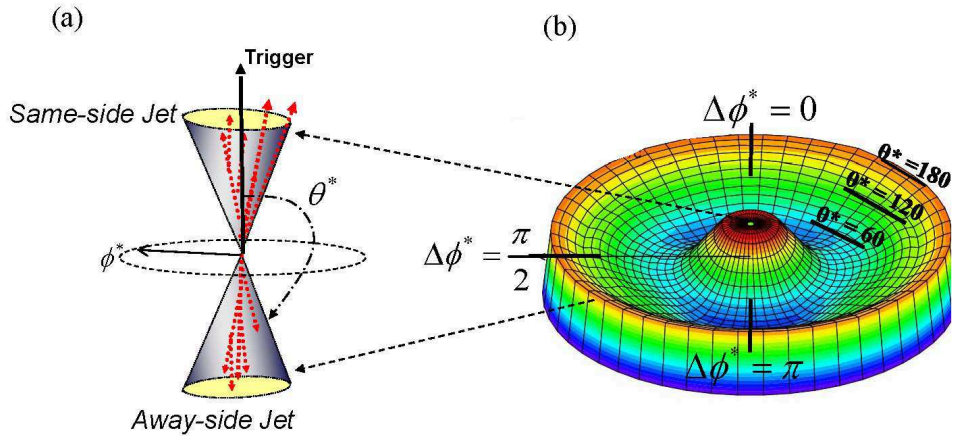}
\caption{(a) Schematic illustration of the 
coordinate system used for the construction of three-particle correlation functions
comprised of high $p_T$ trigger particle and two low $p_T$ associated particles.
The lab angles ($\theta_{lab},\phi_{lab}$) of each particle is transformed  to ($\theta^*,\phi^*$) in a  
new frame whose $z$ axis is the direction of the high $p_T$ trigger. 
(b) Polar plot of a simulated three-particle correlation function for a normal di-jet. 
The polar angle $\theta^*$, of one of the associated low $p_T$ particles in the new frame, is plotted 
along the radial axis; the difference between the azimuthal angles $\Delta\phi^*$, of the two associated 
hadrons in the new frame (see text) is plotted along the azimuthal axis.
}
\label{Fig1}
\end{figure*}

	Analogous to $C_2(\Delta\phi)$, three particle correlation functions can be constructed 
from particle triplets comprised of one high $p_T$ trigger particle (i) and two associated 
particles of lower $p_T$ (j and k). To focus on three-particle jet-like correlations, we transform the 
lab angles ($\theta_{lab},\phi_{lab}$) of each particle to ($\theta^*,\phi^*$) in a  
new frame whose $z$ axis is the direction of the high $p_T$ trigger hadron. This coordinate 
frame is illustrated in Fig.~\ref{Fig1}(a). In this frame, the three particle correlation 
function is given by the ratio of two distributions:
\begin{eqnarray}
C_3(\theta^*,\Delta\phi^*) = \frac{N_S(\theta^*,\Delta\phi^*)}{N_M(\theta^*,\Delta\phi^*)},
\end{eqnarray}
where $N_S(\theta^*,\Delta\phi^*)$ and $N_M(\theta^*,\Delta\phi^*)$ are normalized 
2D-distributions for the same- and mixed-events respectively. Here, 
$\theta^*$ is the polar angle of one of the two associated hadrons and 
$\Delta\phi^*=  \left|\phi_j^*-\phi_k^*\right|$ is the difference between their azimuthal 
angles. The same-events distribution was obtained via event-by-event selection of particle triplets 
from the same event. For mixed-events, particle triplets were obtained by selecting each member 
from a different event.

	The correlation function so obtained is best viewed in 
the polar representation shown in Fig. \ref{Fig1}(b), where a simulated three particle 
correlation surface is shown. Both flow and di-jet correlations were incorporate in 
the simulation. In Fig.~\ref{Fig1}(b), $\theta^*$ and $\Delta\phi^*$ indicate 
the radial and azimuthal axes respectively. In this polar representation, the near-side 
jet is indicated by a peak at the center of the plot ($\theta^*=0^0$) and the characteristic 
ridge at $\theta^*=180^0$ signals a normal or unmodified away-side jet (i.e. a back-to-back jet). 
                                                                                
\begin{figure*}
\includegraphics[scale =0.55]{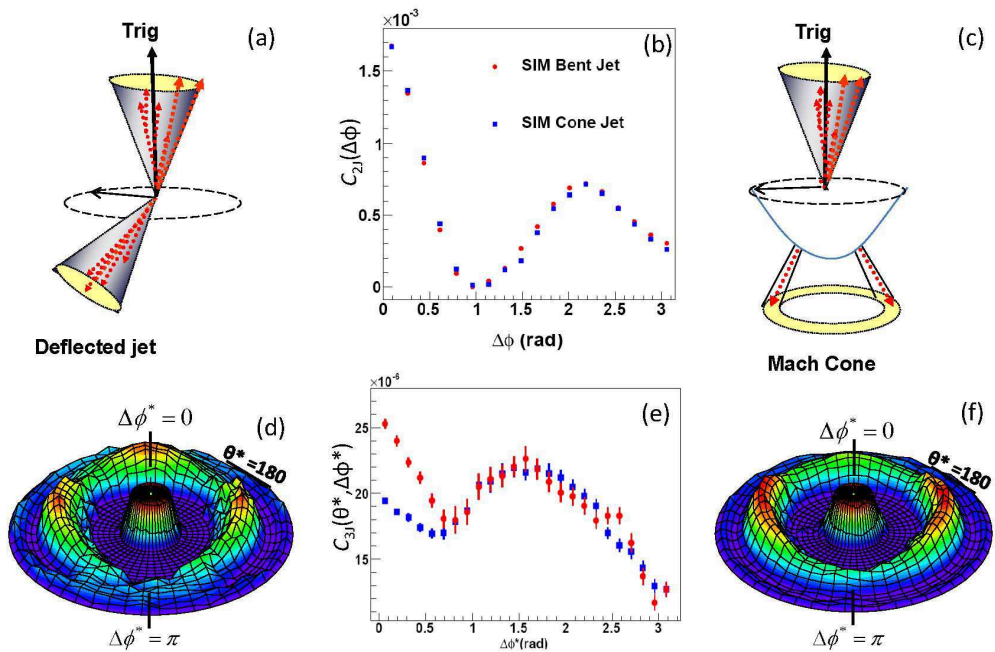}
\caption{ (a) Schematic illustration of a bent jet topology; (b) comparison of the two-particle jet functions 
              for bent jet and conical flow; (c) schematic illustration of conical flow topology; (d) 
              simulated three particle correlation surface for bent jet; (f) simulated three particle correlation 
              surface for conical flow. Panel (e) shows an azimuthal projection of both surfaces for 
              $\theta^* \sim 120^0$ i.e along the ridge (see text).
}
\label{Fig2}
\end{figure*}

	The primary objective of our study is to use such correlation surfaces 
to distinguish between different mechanistic scenarios for away-side jet modification 
which can not be discerned via two-particle correlation functions. 
To demonstrate this ability we used our simulation code to model di-jets 
with (i) an away-side bent jet and (ii) an away-side ``cone'' jet (see illustrations 
in Figs.~\ref{Fig2}(a) and (c) respectively). For the first, 
the away-side jet axis is bent to an angle of $\sim 120^0$ with respect to the 
lead-jet axis and the away-side jet particles are emitted clustered about this axis.
The axis for the away-side cone jet was chosen opposite (on average) to the lead-axis, 
and its associated jet-like particles were emitted so as to mimic a Mach cone with  
Mach angle $\theta_M \sim 60^0$. It is important to note here that these simulations 
were performed for the PHENIX detector acceptance. Equally important is the fact that 
model parameters for the simulations were tuned (with insight from experimental data) to 
give the same shape for the simulated jet-pair correlation functions for bent- and 
cone jets, as shown in Fig. \ref{Fig2}(b). For this simulation set, an isotropic underlying 
event was employed.  

Figure \ref{Fig2}(a) gives an illustration of a di-jet with a bent 
away-side jet. The corresponding simulated three particle correlation 
function is shown in Fig.~\ref{Fig2}(d). This correlation surface exhibits 
a sizable peak at $\theta^* = 0^0$ corresponding to the lead- or same-side jet,
and a ridge at $\theta^* = 120^0$ corresponding to the away-side jet, 
shifted, on average, by $\sim 60^0$ from $\theta^* = 180^0$.  
The projection of $\Delta\phi^*$ for $\theta^* \sim 120^0$ (i.e around the ridge) 
is shown in Fig.~\ref{Fig2}(e); it shows a relatively large peak near 
$\Delta\phi^* = 0^0$, which results from particle triplets comprised of 
a high $p_T$ trigger from the near-side jet and two low 
$p_T$ associated particles from the away-side jet.  Two small peaks  
can also be observed at $\Delta\phi* \sim 90^0\; \text{and}\;270^0$ in Fig.~\ref{Fig2}(d); 
they result from particle triplets in which the high $p_T$ trigger and one low $p_T$ 
particle is from the same-side jet, but the second low $p_T$ particle is from 
the away-side jet. The fall-off at large $\Delta\phi*$ angles reflects the influence 
of the PHENIX detector acceptance as discussed below.

	Figure \ref{Fig2}(f) shows the three particle correlation 
surface for a di-jet with an away-side Mach cone (see illustration 
in Fig. \ref{Fig2}(c)). Analogous to Fig.~\ref{Fig2}(d), a near-side peak 
at $\theta^*  = 0^0$, and an away-side ridge near $\theta^*  = 120^0$ is 
apparent. However, in contrast to the case for the bent jet, the   
peak at or near $\Delta\phi^*  = 0^0$ is less pronounced. This distinctive 
signal is made more transparent in Fig. \ref{Fig2}(e) where the azimuthal 
projections ($\Delta\phi^*$ along the ridge) for the away-side bent jet and 
Mach cone are compared. These correlations result from particle triplets comprised 
of a high $p_T$ trigger from the same-side jet and two low $p_T$ particles 
from the away-side Mach cone. As in the case of the bent jet, 
the two small peaks at $\Delta\phi* \sim 90^0\; \text{and}\;270^0$ in 
Fig. \ref{Fig2}(f), reflects ``anomalous'' particle triplets in which the high $p_T$ 
trigger and one low $p_T$ particle is from the same-side jet and the 
second low $p_T$ particle is from the away-side Mach cone. 

	Intuitively, correlated triplets which result from a near-side high $p_T$ trigger 
and two associated particles from an away-side Mach cone, should lead to an 
away-side azimuthal ridge at $\theta^* = 180^0 - \theta_M$ with no preference 
for $\Delta\phi^*$. Therefore, the correlation surface in Fig. \ref{Fig2}(f), as 
well as its $\Delta\phi^*$ projection in Fig. \ref{Fig2}(e) gives an indication of the 
relative influence of the PHENIX detector acceptance (especially in the 
vicinity of $\phi^*$ angles close to $180^0$) and the anomalous triplet correlations 
discussed above.

\subsection{\label{sec:level1}Suppression of harmonic correlations}

The correlation functions shown in Fig.~\ref{Fig2} were obtained from 
simulated events with hits recorded in the PHENIX acceptance from three 
di-jet-like particles, one trigger (high $p_T$) and two associated (lower 
$p_T$) particles; the underlying event particles were made isotropic.  It is well 
known however, that there are strong harmonic correlations between the various 
particles and the reaction plane that follow Eq.~\ref{v2v4}.  Therefore, the 
underlying event is not actually isotropic. Detailed systematic measurements have 
been made of the Fourier coefficients $v_{2,4}$ with a significantly reduced 
influence from jet-like correlations \cite{ppg098,Lacey:2009xx}; 
thus, it is straightforward to incorporate these harmonic correlations into 
the event simulations.  

\begin{figure*}
\includegraphics[scale =0.6]{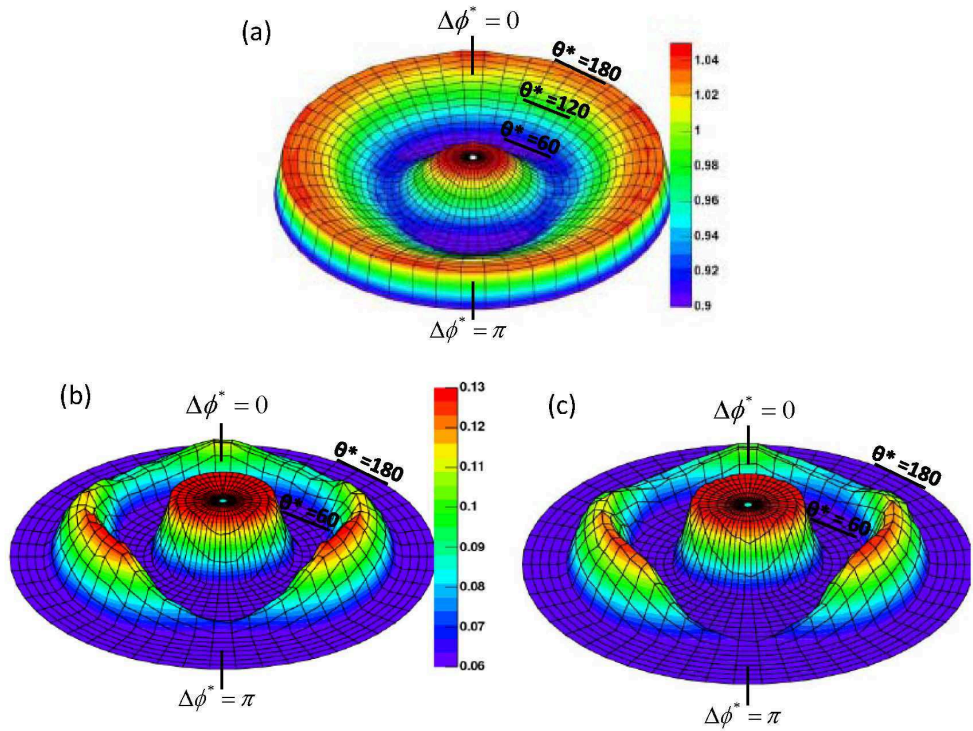}
\caption{ (a) Simulated three particle correlation surfaces; the simulation 
          included both flow and jet-like correlations. (b) Same as (a) but 
          after flow subtraction. (c) Input jet-like correlations only. 
          Note the similarity between (b) and (c).
}
\label{Fig3}
\end{figure*}

A three particle correlation surface which results from the combined influence 
of jet-like and flow correlations is shown in Fig.~\ref{Fig3}(a). It shows that the 
combined correlations make it more difficult to visualize the detailed effects of 
the di-jet-like emissions in the raw correlation function. This is akin to the actual 
experimental situation. Therefore, we wish to remove the 
harmonic correlation effects from the three-particle correlation 
surfaces. To this end we follow a procedure similar to that described in 
reference \cite{Ajitanand:2005jj} using the ZYAM assumption i.e. we assume that 
the jet correlation function has zero yield at its minimum.

The three-particle correlation function which results from a combination of  
flow and jet-like sources can be given as
\begin{eqnarray}
C_3(\theta^*,\Delta\phi^*) = a_0\left[C_H(\theta^*,\Delta\phi^*)+C_J(\theta^*,\Delta\phi^*)\right].
\end{eqnarray}
The harmonic flow correlation $C_{H}(\theta^*,\Delta\phi^*)$ is obtained from 
a flow only simulation where  particles are emitted according to the measured
pattern with respect to the reaction plane.
The jet correlation is then given by 
\begin{eqnarray}
C_{J}(\theta^*,\Delta\phi^*) =\frac{C_3(\theta^*,\Delta\phi^*)-a_0C_{H}(\theta^*,\Delta\phi^*)}{a_0}.
\end{eqnarray}
Applying the ZYAM condition one obtains 
\begin{eqnarray}
a_0C_{H}(\theta^*_{min},\Delta\phi^*_{min}) = 
C_3(\theta^*_{min},\Delta\phi^*_{min})
\end{eqnarray}
which can be solved to obtain $a_o$. 

Tests for flow removal for the case of three-particle correlation 
functions are shown in Fig. \ref{Fig3}.  
Fig.~\ref{Fig3}(a) shows the correlation surface obtained from a 
simulation that includes both di-jet and harmonic correlations.  
Fig.~\ref{Fig3}(b) shows the jet function that results after removal of 
the harmonic correlation by the ZYAM-driven subtraction.  This is to be 
compared to the input jet-like correlation function generated with no harmonic 
correlations, shown in Fig.~\ref{Fig3}(c). The similarity 
between  Figs.~\ref{Fig3}(b) and (c) clearly show that the ZYAM procedure 
is able to successfully recover the essential characteristics of the 
input di-jet-like function.

\subsection{\label{sec:level1}Removal of (2+1) correlations}

	The three particle correlation function $C_J(\theta^*,\Delta\phi^*)$, obtained after 
removal of flow effects, still contain contributions from false triplets or (2+1)-correlations.
They are of two types: (i) False ``hard-soft'' ($hs$) triplets 
in which the high $p_T$ trigger and one associated low $p_T$ particle  
come from the di-jet, but the  second low $p_T$ associated particle is from the underlying event. 
(ii) False ``soft-soft'' ($ss$) triplets in which the two low $p_T$ associated particles belong 
to the same di-jet but the high $p_T$ trigger is from the underlying event. 
An estimate of such correlations can be made and removed as follows. 
First, a (2+1)-correlation function 
\begin{eqnarray}
C^{(hs+ss)}_{(2+1)}(\theta^*,\Delta\phi^*) = \frac{N_{S(2+1)}(\theta^*,\Delta\phi^*)}{N_M(\theta^*,\Delta\phi^*)},
\end{eqnarray}
is constructed from event pairs. Here, 
the distribution for fake-triplets $N_{S(2+1)}(\theta^*,\Delta\phi^*)$ is obtained 
by taking two particles from one event ($hs$ or $ss$) and a third from another event. 
The $hs$ and $ss$ triplets are sampled in the ratio of observed $ss$
and $hs$ two-particle correlation strengths. As before, the mixed-events distribution 
$N_M(\theta^*,\Delta\phi^*)$ is obtained by taking each member of a particle triplet 
from a different event. This correlation function estimates both the $ss$ 
and $hs$ components of the (2+1)-contribution.
Second, the flow contribution to $C^{(hs+ss)}_{(2+1)}(\theta^*,\Delta\phi^*)$ was subtracted 
via the procedure outlined earlier, to obtain the (2+1)-jet-like contribution $C_{J(2+1)}(\theta^*,\Delta\phi^*)$.  
The fully corrected triplet correlation function $C_{3J}(\theta^*,\Delta\phi^*)$ was 
obtained via flow and (2+1)-subtraction;
\begin{eqnarray}
C_{3J}(\theta^*,\Delta\phi^*) = C_3(\theta^*,\Delta\phi^*) - H(\theta^*,\Delta\phi^*) \nonumber \\ 
       - C_{J(2+1)}(\theta^*,\Delta\phi^*).
\end{eqnarray}

	We have made extensive tests of this analysis procedure via detailed simulations 
and have confirmed its utility in suppressing both flow and (2+1)-correlations.
\begin{figure*}
\includegraphics[scale =0.7]{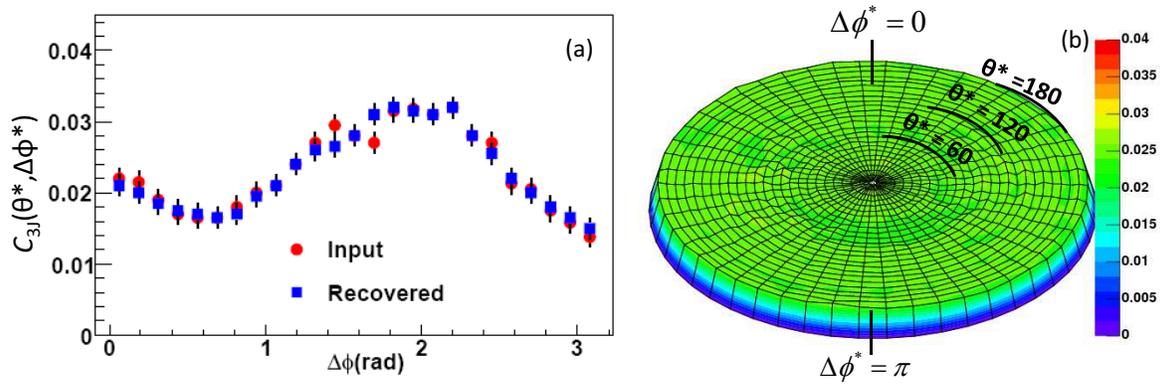}
\caption{(a) Comparison of the $\phi^*$ projections for $\theta^* \sim 120^0$ for simulated input 
             and output correlation functions. The output correlation function is obtained after 
             harmonic and (2+1) subtraction (see text); (b) correlation surface obtained from 
             a simulation in which only (2+1)-correlations are introduced i.e no genuine three 
             particle correlations are introduced. A small offset is added for clarity. 
}
\label{Fig4}
\end{figure*}
Fig.~\ref{Fig4}(a) shows a representative comparison of the $\Delta\phi^*$ projections [along the ridge] 
for an input (filled circles) and a recovered (filled squares) 
three-particle correlation function. The recovered correlation function is obtained after 
removing the effects of flow and (2+1)-contributions.  
The comparison clearly speaks to the efficacy of the technique.
A further confirmation of the efficacy of our procedure can be illustrated via 
the correlation surfaces obtained from simulations which include only (2+1)-correlations. 
One such example is given in Fig.~\ref{Fig4}(b); it shows the expected ``flat'' and featureless 
surface that is to be expected from the successful removal of (2+1)-contributions.

\section{\label{sec:level1} Summary}

Detailed experimental probes of the hot and dense medium created in 
ultra-relativistic heavy ion collisions, are currently being explored 
via measurements of multi-hadron correlations at RHIC. These correlations 
contain contributions from both harmonic flow and jet-like emissions which must be 
disentangled, as well as topological features important to the study of the mechanism 
for modification of the away-side jet. A method is presented for the analysis of such 
data to retrieve strongly modified jet-like topologies via three particle 
correlation functions. In particular, we have shown that 
a distinction between conical flow and a deflected (or bent) jet 
can be obtained in our analysis framework.
Intuitively, one knows that a normal or back-to-back jet would generate 
a peak for the near-side jet centered at $\theta^* = 0^0$ and an azimuthal ridge at $\theta^* 
= 180^0$.  Further, it is expected that a bent jet would lead to an away-side 
azimuthal ridge shifted to a $\theta^*$ angle less than $180^0$, but 
would cluster the associated away-side jet-like particles near to $\Delta\phi^* = 0^0$.  
By contrast, a Mach cone should lead to an away-side azimuthal 
ridge also shifted to a $\theta^*$ angle less than $180^0$, but without preference 
for any $\Delta\phi^*$ value. For the most part, these are the patterns shown by the 
simulations, albeit with some distortions due to the limited $\eta$ acceptance 
of the detector and the detection of anomalous triplets. 
However, these distortions do not blur the topological 
distinction between an away-side bent jet and conical flow.

\bibliography{3pMeth}

\end{document}